\documentclass[conference]{IEEEtran}
\usepackage[latin1]{inputenc}
\usepackage{amsmath}
\usepackage{amsfonts}
\usepackage{amssymb}
\usepackage{graphicx}
\usepackage{multicol}
\usepackage[usenames]{color}
\usepackage{url}

\begin{document}
%
\title{A Survey on Multipath Transport Protocols}

\author{Bachir Chihani - Denis Collange\\
Orange Labs  Sophia Antipolis, France\\
Email: {bachir.chihani,denis.collange}@orange-ftgroup.com\\
}

\maketitle


\begin{abstract} 
We present a survey on multipath transport protocols. These protocols are aiming to provide a way for the use of simultaneous paths at the transport layer and load balancing traffic on these paths. We describe some of the main proposal and then we focus on MPTCP (Multipath TCP) which is a promising extension of TCP currently considered by the recent eponymous IETF working group. 
\end{abstract} 


\section{Introduction} 
Nowadays mobile equipment have often more than one single network interface. For instance, laptops have usually at least both a wired (Ethernet) and a wireless (Wifi) network adapters. Similarly smartphones and tablet PCs can reach the Internet either through Wifi or through a cellular network (UMTS or 3G+). 

Another fact is that operators usually duplicate links and network equipments to protect their networks against failures, especially in their access and backhaul networks. 
Moreover the backbone networks are generally meshed. 
In this context many paths can exist between any two endpoints. The idea to use concurrently many paths has then emerged, in order to improve the robustness and performance of end-to-end connections. Such multipath connections can indeed balance the load between the different paths, switch dynamically the traffic to the best path, avoiding congested or broken links. 

A lot of studies have considered the implementation of multipath capabilities at different layers: at the application layer \cite{hacker02}, at the transport layer \cite{ong02}, \cite{hasegawa05}, \cite{dong07}, \cite{sarkar06}, \cite{zhang04}, \cite{rojviboonchai02}, \cite{hasegawa05}, etc.

We think that it is at the transport layer where end-systems can make maximum benefit of the multipath \cite{wischik08}. At this layer, end-systems can gather information about each used path: capacity, latency, congestion state. These information can then be used to react to congestion in the network by moving the traffic away from congested paths. 

An IETF's working group has recently been created to standardize a multipath protocol at the transport layer. They proposed Multipath TCP \cite{ford10} (MPTCP), an extension of TCP to handle multiple paths between two endpoints. 

The reminder of this paper is as follow: we first present in section \ref{sec:multipath} some protocols handling multipath at the transport layer. Then we describe MPTCP in section \ref{sec:drafts}, as it is specified in the current versions of the IETF drafts. In section \ref{sec:discussion}, we discuss the implemented mechanisms in the different cited protocols. Finally, we conclude the paper in section \ref{sec:conclusion}. 

\section{Multipath transport protocols}\label{sec:multipath}
\subsection{ATLB (Arrival-Time matching Load-Balancing)}
ATLB \cite{hasegawa05} is a transport protocol supporting the multipath. It allows the distribution of data among different available paths with the objectives to minimize the dis-sequencing of packets at the receiver side, the detection of problems on paths, and the recovery of lost packets.
ATLB provides a way to mesure the arrival time of packet to the receiver. It defines this time as the queuing delay plus the network delay (smoothed Round-Trip-Time). 
Using these delays, ATLB attributes a score for each path and use the path with the least score to send data.

\[ score_{i} = \frac{Q_{i}}{G_{i}} + \frac{sRTT_{i}}{2} \]
$Q$ is the length of data in the sending buffer; $sRTT$ is the smoothed Round-Trip-Time; $G$ is a smoothed throughput computed as $G_{j} = \alpha * G_{j}-1 + (1 - \alpha) * TPUT_{j}$. $\alpha (0 < \alpha < 1)$ is a constant, and $TPUT_j$ is the throughput of a TCP connection measured each $\beta$ milliseconds.

For path failure detection, ATLB maintains a timer which expires after a time $T$. ALTB assumes that a lost segment after an RTO expiration means a path failure or a path highly congested. Thus $T = RTO + \theta * RTT ( \theta > 2)$.

In case of path failure, ATLB sends a sensing packet each $P$ ms. It also compute the lost rate each $S$ ms. If the lost rate is less than $R$\%, then it assume that the packet is recovered and can be used.

\subsection{Fair-TCP}
Fair-TCP \cite{kancherla07} is a protocol supporting the multipath, implemented at both the sender and the receiver side. It was conceived for SAN (Storage Area Network) where the TCP sessions are maintained for a long period of time and the exchanges are based on SCSI input/output commands, and which are part of a communication standard called iSCSI (Internet SCSI). 
For enhancing the performance of the communications, Fair-TCP shares congestion information between the different connections by the use of a data structure called the ECB (Ensemble Control Block) which is an extension of the TCP Control Block.

\subsection{PATTHEL}
PATTHEL \cite{baldini09} is a solution implemented at the session layer to paralelly transfert data. It provides an API (Application Programming Interface) for the application developper to use this protocol. 
The protocol uses a dedicated channel (end-to-end path) created in first for controlling the connection, the rest channels are used to transfert data. 
A received data block from the application layer is divided into chunks of variable size depend of the channel characteristics. To each chunk a header is added which contain:
\begin{itemize}
\item Chunk size (32 bits) to indicate the size of the payload;
\item Stream offset (64 bits) to indicate the position of the data on the flow;
\item Block size (32 bits) allows the receiver to check if the block can be hold in the application buffer;
\item Block index (32 bits) used to check if the chunk belong to the block currently in reception.
\end{itemize}

To force sending packets over a certain channel, PATTHEL add an entrance to the routing table. 

\subsection{R-MTP (Reliable Multiplexing Transport Protocol)}
R-MTP \cite{magalhaes01} is to used by mobile nodes having many wireless interfaces of potentially heterogenous technologies.
It is a transport protocol able to agregate the available bandwidth of different network paths by distributing data over these paths. The protocol maintain a set of information about each used path in order to react to any change happening over a path. Like the bandwidth which is estimated by the Packet Pair method. This one helps to estimate the least time interval between packets so that to avoid queuing delays.
The protocol makes use of a special header format to exchange information between endpoints and SACK (Selective Acknowledgements) for reliability.
Example of the exchanged information which can be included in the header:
\begin{itemize}
\item \textit{Initial rate} is the reverse of the minimum period (between sending two packets) that the path can support without creating congestion;
\item \textit{Interarrival time} is the difference between the arrival time of the precedent packet and the current one, on the same path;
\item \textit{Jitter} is the difference between the measured interarrival time and the rate on the same path;
\item \textit{Commulative long run jitter} is the commulated jitter which in case it grows can be interpreted as a congestion.
\end{itemize}

\subsection{cTCP (Concurrent TCP)}
cTCP \cite{dong07} is a TCP-based protocol which allows the use of multiple paths between two hosts having many network interfaces. Figure \ref{cTCP} illustrates the architecture of this protocol which is composed of a packet scheduler used for laod-balacing the traffic on the different paths; an acknowledgment processor used to fix the gap report problem in the TCP congestion control. This component include a Duplicated ACK classifier which can provide information  about the quality of a path to the packet scheduler; An interne databases implemented as a chained linear list.
To remain compatible with TCP standards versions, cTCP uses a single congestion window to control the global throughput and a single emission buffer shared between the different paths.
At the connection establishment, the two enpoints exchange their available addresses using specific option, if one of the endpoint isn't a cTCP one it will ignore the options putting the connection to be a standard TCP one.
The used packet scheduling algorithm is a Credit-Weighted Round-Robin. This one allows a fair data distribution among the different paths. Whanever an acknowledgement is received, the estimated bandwidth of the corresponding path is updated and a new sending credit is added to the sender. The new credit is then divided between the different paths.

For the congestion control, cTCP uses the database to store all unacked packet with the identifier of the path used to send the packet. When a new ackowledgement is received, the sender drop all packet having a sequence number less the one included in the acknowledgment. When a duplicated acknowledgment is received, the sender checks if the path used to send the packet suspected to be lost and the path from which the dupack has been recevied are the same. If it is, then a Fast Retransmit occurs, otherwise the dupack is ignored.

\section{Multipath TCP}\label{sec:drafts}
An IETF's working group has been created to standardize a multipath protocol for the transport layer. They proposed MPTCP \cite{ford10} (Multipath TCP), an extension of TCP to handle multiple paths  between two endpoints. MPTCP is designed with three major goals:

\begin{enumerate}
\item 	{\bf Improve throughput}: the performance of a multi-path flow should be at least as good as this of a single-path flow on the best route.
\item 	{\bf Do no harm}: a multi-path flow should not take up any more capacity on any one of its paths than a single-path flow using that route.
\item 	{\bf Balance congestion}: a multi-path flow should move as much traffic as possible away from the most congested paths.
\end{enumerate}

\subsection{Main mechanisms}
With MPTCP, the transport layer is splitted in two sublayers. The upper one gathers the functionalities for connection management (establishing connection, reordering packets, etc.). The lower one gathers a set of subflows that can be seen as one TCP flow.
MPTCP distinguishes two spaces for sequence numbers. Each subfow has its own sequence space which is similar to the Standard TCP sequence number, identifying bytes within a subflow. At the connection level, another sequence space is used for reordering purposes.

The MPTCP protocol use new TCP options to exchange signalling information between peers, for instance: 
\begin{description}
\item[{\bf MPC}] (Multipath Capable) is used during the three-way handshake to establish a multipath TCP connection.  
\item[{\bf DATA}] {\bf FIN} is used to inform the remote peer of the end of data and to close the multipath TCP connection.
\item[{\bf ADD}] and {\bf REMOVE} Address (Ipv4) are used to inform the remote peer of the availability of a new address or to ask it to ignore an existing one.
\item[{\bf JOIN}] is used to initiate a new sub-flow (packet flow on a route) between a not used peer of addresses.
\item[{\bf DSN}] (Data Sequence Number) is used as a map between subflow level and data sequence space number.
\end{description}

\twocolumn

\subsubsection{Connection establishment}
\begin{figure}[h]
\centering
\includegraphics[scale=0.5]{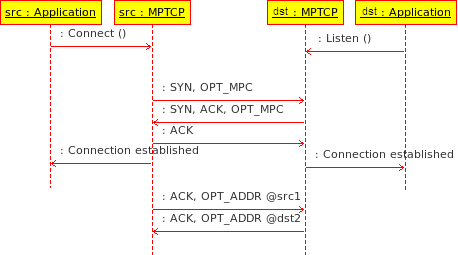} 
\caption{MPTCP connection establishment}
\label{ConEstab}
\end{figure}

Figure \ref{ConEstab} illustrates the process of establishment of a MPTCP connection. After that the source application sends a Connect() call, the transport layer establishes a connection with the destination peer which was waiting for receiving connection requests. The establishment is TCP-like (three way handshake) with the use of MPC option to inform the other peer that the initiator is able to exchange data using multipath TCP. To initiate subflows, peers must first exchange their additional addresses. The MPCTP draft do not specify how the exchange may happen. We choosed to send additional TCP segments. These segments handle the ADDR (Add address) option and are sent just after successfully establishing the connection.

\subsubsection{Subflow initiation}
\begin{figure}[h]
\centering
\includegraphics[scale=0.5]{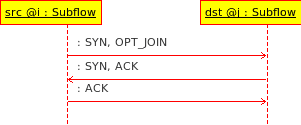} 
\caption{MPTCP sub-flow initiation}
\label{SubflowInit}
\end{figure}

Figure \ref{SubflowInit} shows the initiation of a new subflow and the presence of a JOIN in a SYN segment. To maximize the chance that the subflow under initiation takes a path which is disjoined with previously established paths, an address is used only by a one subflow.

\subsection{Traffic control}
MPTCP redefines some TCP mechanisms so that they fit the multi-path context. 
Congestion control allows the sender to regulate its throughput according to the available network resources. 

In MPTCP, the congestion control is performed at the subflow level. Each subflow has its own congestion window. But the congestion windows of different subflows may be coupled to improve the performance. 

In MPTCP, the receiver has only one global receiving window shared between the set of the established subflows. The objective is to do not limit the speed of some subflows.

Four different algorithms have been proposed by Raiciu et al. in \cite{raiciu09}, coupling in various ways the congestion windows of active subflows: Uncoupled, Fully Coupled, Linked Increase, and RTT Compensator. They consider a simple extension of TCP's congestion control in case where the round-trip time is the same for all the available paths $r = 1,..., N$. 

With the algorithm Uncoupled, the congestion window of each subflow behaves like for a single Standard TCP connection. 
Let $w_r$ be the congestion window on path $r$, and $w = \sum_r w_r$

Algorithm {\bf Fully Coupled}
\begin{itemize}
\item $w_r = w_r + \frac{1}{w}$ per ACK on path $r$
\item $w_r = max(w_r - \frac{w}{2}, 1)$ per loss event on path $r$
\end{itemize}

Most of the time either one path or another is used with this algorithm, and rarely both. This phenomenon is called Flappiness. To reduce flappiness, authors proposed the following algorithm:

Algorithm {\bf Linked Increases} \cite{raiciu10}
\begin{itemize}
\item $w_r = w_r + \frac{a}{w}$ per ACK on path $r$
\item $w_r = \frac{w_r}{2}$ per loss event on path $r$
\end{itemize}

In more general case where RTT (round-trip time) are not equal for the all paths, the authors adjust the precedent algorithm:

Algorithm {\bf RTT Compensator}
\begin{itemize}
\item $w_r = w_r + min(\frac{a}{w}, \frac{1}{w})$ per ACK on path $r$
\item $w_r = \frac{w_r}{2}$ per loss event on path $r$
\end{itemize}

\section{Discussion}\label{sec:discussion}
Follow we describe the different mechanisms that a reliable and connection oriented transport protocol msut have them, also the modification which must occurs to these mechanisms for adapting them to the multipah.

\subsection{Congestion control}
For a multipath TCP connection, the congestion control must happen in each used paths so that we can match in real-time manner the quantity of data sent on a path and the capacity of this one. Also, a congestion control for a multipath TCP solution must satisfy the previously stated goals: \textit{improve throughput, do no harm, balance congestion}.
Like in TCP and for each path, the sender has to maintain a set of parameters for the congestion control and the updates after receiving acknowledgement and duplicated acknowledgement: RTT, RTO, ssthresh, cwnd, etc.
The congestion control must happen on each path in takking into account its characteristics. It is foreseeable that a coupling strategy can be used to combin the differents paths parameters in order to aggregate the ressource of the available paths, or move away data from congested paths.

\subsection{Flow control}
Most of the proposed solutions for flow control in the multipath transport protocols are a simple adaptation of the TCP control flow. The receiver maintains a single window shared between all subflows in order to not limit the speed of some of them. The sender and the receiver agree on the awnd (advertised window) which represents the quantity of data which can be sent without exchanging acknowledgement. The sender can then divides the window size between all the available paths according to the used data distribution policy.

\subsection{Acknowledgement management}
When receiving an acknowledgement on a path, the corresponding parameters are updated (round trip time, bandwidth, loss rate, congestion window). If some data need to be retransmitted, it is possible to do it on the same path used firstly to transmitted these data (even if that path is potentially congested), or on another one which is the preferable solution because the retransmission will be faster. 

\subsection{Loss packets recovery}
In standard TCP, loss recovery happens after an RTO (Retransmission TimeOut) expiration, or the reception of three duplicated acknowledgment. The recovery consists of sending all not acked data. In a multipath context, the same mechanisms can be reused but after adaptation. For the duplicated acknowledgement we can vary the threshold, differentiate between spurious acknowledgements caused by the arrival of packets in out of sequence manner at the receiver, and the real acknowledgement stating a potential lost. For the RTO, for each used path needs a specific RTO so that when the RTO expires only the concerned path will be affected.

\subsection{Failure management}
To make sure that a path which wasn't used for a certain time is back operationnal, the protocol may use sensing message (like Heartbeat in SCTP) in defined interval. If all  paths are used during a communication, the protocol may use a counter returned to zero each time some data are received and in case of expiration the path can be considered as fail.

\subsection{Data distribution over paths}
Most of the proposals use a Round Robin based sequencing policy, and distribute data fairly distributed. But paths have characteristics (latency, capacity, jitter, loss rate, etc.) which are potentially differents, a such policy is not intersting for multipath. 
Other protocols use an ameliorated Round Robin policy and distribute on each path an amount of data which is proportional to the throughput of the used path, ou only send data on the best path (like ATLB, M/TCP). Another proposal consist of sending data out of sequence and with proportional amount to the path characteristics so that data arrives in sequence to the receiver. A distribution method can take into account any combinaison of the parameters capacity, latency, jitter, loss rate, etc. But it has to ensure that data arrive in sequence without an overload a path while others are not used.

\subsection{Managing out of sequence data arrival}
Packets may take different paths, thus they may arrive at the receiver out of sequence. The receiver has to hold a free space to save these packets from the other which are makin a normal sequencing. 

\subsection{Path managment}
Path management (i.e. adding or droping paths) may be done by using the option field of the TCP header. For choosing paths, these have to be disjoined which means they do not share the same physical link. Otherwise, in case of congestion or failure of one of them, all others will also be congested or fail. For the connection management, most of the presented protocols, even those no compatible to TCP, use the TCP's three way handshake.

\section{Conclusion}\label{sec:conclusion}


\begin{thebibliography}{1}
  
\bibitem{dong07} 
Yo Dong, N. Pissinou, and J. Wang. "Concurrency Handling in TCP." 5ème conférence annuelle en Communication Networks and Services Research (CNSR'07). 2007 IEEE.

\bibitem{kancherla07} 
B. Kancherla, G. Narayan, K. Gopinath. "Performance Evaluation of Multiple TCP connections in iSCSI", 24th IEEE Conference on Mass Storage Systems and Technologies (MSST '07), 2007. 

\bibitem{magalhaes01} 
L. Magalhaes and R. Kravets. "Transport level mechanisms for bandwidth aggregation on mobile hosts". ICNP (2001), 0165. 

\bibitem{baldini09} 
A. Baldini, L. De Carli et F. Risso. "Increasing Performance of TCP Data Transfers Through Multiple Parallel Connections", IEEE Symposium on Computers and Communications (ISCC 09),  Sousse, Tunisia,  July 2009.

\bibitem{hacker02} 
T. Hacker and B. Athey, "The End-to-End Performance Effects of Parallel TCP Sockets on a Lossy Wide-Area Network ," in IEEE IPDPS, Florida, Apr. 2002. 

\bibitem{ong02} 
L. Ong, J. Yoakum, "An Introduction to the Stream Control Transmission Protocol (SCTP)," RFC 3286, May 2002. 

\bibitem{hasegawa05} 
Y. Hasegawa, I. Yamaguchi, T. Hama, H. Shimonishi and T. Murase. "Improved Data Distribution for Multipath TCP communication," IEEE Globecom 2005. 

\bibitem{wischik08} 
D. Wischik, M. Handley and M. Bagnulo Braun, "The Resource Pooling Principle", ACM SIGCOMM Computer Communication Review, Vol 38.5, pp 47-52, October 2008. 

\bibitem{ford10} 
A. Ford, C. Raiciu and M. Handley, "TCP Extensions for Multipath Operation with Multiple Addresses," draft-ietf-mptcp-multiaddressed-01, July 12, 2010. 

\bibitem{raiciu09} 
C. Raiciu, D. Wischik and M. Handley, "Practical Congestion Control for Multipath Transport Protocols," UCL Technical Report, 2009. 

\bibitem{raiciu10} 
C. Raiciu, M. Handley and D. Wischik, "Coupled Multipath-Aware Congestion Control", draft-raiciu-mptcp-congestion-01, March 2010. 

\bibitem{sarkar06} 
D. Sarkar."A Concurrent Multipath TCP and Its Markov Model," IEEE ICC proceedings, 2006. 

\bibitem{zhang04} 
M. Zhang, J. Lai, A. Krishnamurthy, L. Peterson and R. Wang. "A transport layer approach for improving end-to-end performance and robustness using redundant paths," In ATEC'04: Proceedings of the annual conference on USENIX Annual Technical Conference (Berkeley, CA, USA, 2004), USENIX Association, pp. 8-8. 

\bibitem{rojviboonchai02} 
K. Rojviboonchai and H. Aida. "An evaluation of multi-path Transmission Control Protocol (M/TCP) with robust acknowledgement schemes," Proc. Internet Conference 2002 (IC2002), Univ. Tokyo, Japan, October 2002. 

\end{thebibliography}
\end{document}